# Spatial and Temporal Correlates of Vesicular Release at Hippocampal Synapses

Suhita Nadkarni[1, 2, *], Thomas Bartol[1, 2, *], Terrence Sejnowski[1, 2, 3] Herbert Levine[1]

*1. Center for Theoretical Biological Physics, University of California at San Diego, 9500 Gilman Drive, La Jolla Ca 92093,USA*

*2. Howard Hughes Medical Institute, Salk Institute for Biological Studies, 10010 North Torrey Pines Road, La Jolla 92037 USA*

*3. The Division of Biological Sciences, University of California at San Diego, 9500 Gilman Drive, La Jolla Ca 92093,USA.*

**. These authors contributed equally to this work*

## Author Summary

Neurons communicate at chemical synapses, where neurotransmitter released from a nerve terminal of the presynaptic neuron signals to the postsynaptic neuron that an event has occurred. The release is triggered by the entry of calcium ions into the nerve terminal. Previously the chemical reactions underlying neurotransmitter release were studied in a giant nerve terminal with many release sites. The goal of our research was to model the release at a much smaller synapse found in the hippocampus, a part of the brain that is involved with learning and memory. The synapse model was simulated in a computer that kept track of all of the important molecules in the nerve terminal so the results can be directly compared with experimental data. The model led to a better understanding of the conditions that lead to the release of a single packet of neurotransmitter, called a quantum. According to our model, the release of more than one quantum can be triggered by a single presynaptic event but the quanta are released one at a time. Furthermore, we uncovered the mechanisms underlying an extremely fast form of release that had not been previously studied. The model made predictions for other properties of the synapse that can be tested experimentally. A better understanding of how the normal synapses in the hippocampus work will help us to better understand what goes wrong with synapses in mental disorders such as depression and schizophrenia.

## Abstract

We develop a spatially explicit biophysical model of the hippocampal CA3-CA1 presynaptic bouton to study local calcium dynamics leading to vesicle fusion. A kinetic model with two calcium sensors is formulated specifically for the CA3-CA1 synapse. The model includes a sensor for fast synchronous release that lasts a few tens of milliseconds and a sensor for slow asynchronous release that lasts a few hundred milliseconds. We show that a variety of extant data on CA3-CA1 synapse can be accounted for consistently only when a refractory period of the order of few milliseconds between releases is introduced. Including a second sensor for asynchronous release that has a slow unbinding site and therefore an embedded long memory, is shown to play a role in short-term plasticity by facilitating release. For synchronous release mediated by Synaptotagmin II a third time scale is revealed in addition to the fast and slow release. This third time scale corresponds to 'stimulus -correlated super-fast' neurotransmitter release. Our detailed spatial simulation indicates that all three-time scales of neurotransmitter release are an emergent property of the calcium sensor and independent of synaptic ultrastructure. Furthermore, it allows us to identify features of synaptic transmission that are universal and those that are modulated by structure.

## Introduction

The synapses from Schaffer collaterals of CA3 pyramidal cells onto CA1 neurons have been extensively studied as sites for learning and memory. Most of these synapses have one or two active zones, thereby allowing easy quantification of vesicular release [1,2,3]. However, due to its small size it is not yet feasible to carry out quantitative local [$Ca^{2+}$] measurements at these synapses, and there exists no kinetic description that can relate calcium dynamics to neurotransmitter release and to excitation history.

In contrast, the calyx of Held is a giant pre-synaptic terminal with hundreds of active zones, that can be probed directly because of its large size [4,5]. However, these active zones are separated from points of calcium entry (i.e. voltage-dependent calcium channels) over a range of distances. This makes it difficult to disentangle the properties of vesicular release that arise due to the kinetics of the calcium sensors alone from those due to their spatial arrangement. Elegant calcium-uncaging experiments have been performed to ensure a uniform calcium concentration across the hundreds of docked vesicles [6,7]. However, the calcium concentration stays high for a long time in these protocols, depleting the docked vesicle resources and modifying the average vesicle release rates. Furthermore, uncertainties in actual number of docked vesicles introduce error in the kinetic models. These difficulties have led to disparate models with calcium sensitivities that vary over 500% [6,7]. For example Fig. 1 in [8]shows that 25 % release probability corresponds to peak calcium of either 8.8 μM or ~50 μM in two separate kinetic models for the calyx. These models provide a starting point but cannot be directly used to provide an accurate description of release at CA3-CA1.

Here we construct a computational spatially explicit model to realistically simulate the neurotransmitter release dynamics at the CA3-CA1 hippocampal synapse. Our model relies on known ultrastructural details such as the average bouton size (~0.5 micron wide), the number of active zones (typically one) and the number of docked vesicles at each active zone (~7) [1]. The overall logic of this study is to draw on the common features of release at central synapses (calcium sensor, multiple time scales, calcium channel kinetics and clustering, buffer kinetics) including all the data available for the calyx of Held and CA3-CA1 synapse. The resulting model for release at the CA3-CA1 synapse can consistently explain these data and leads to a better understanding of the mechanisms underlying synaptic transmission.

**CA3-CA1 Model**

We simulated the sequence of events at the CA3-CA1 synapse beginning with the arrival of an action potential, the opening of P/Q type voltage dependent calcium channels (VDCCs), diffusion of calcium from the VDCC's to the calcium sensor and the triggering of vesicle fusion and glutamate release [9]. The dynamics of these events were simulated stochastically in 3D using Monte Carlo methods (MCell version 3). The canonical CA3-CA1 en passant synapse geometry used in our simulations is shown in Fig. 1A. The model consists of 1) a cluster of voltage-dependent calcium channels (VDCCs) of type P/Q [10] (Fig. S1A), the main contributor to presynaptic $Ca^{2+}$ current in mature hippocampal presynaptic terminals [11,12] 2) plasma membrane calcium ATPase (PMCA) pumps that work to keep the base level $Ca^{2+}$ at 100 nM (Fig. S1B) 3) the mobile calcium buffer calbindin-D28k [13] (Fig. S1C) 4) an active zone populated by seven docked vesicles [1,14], each endowed with its own calcium sensor for neurotransmitter release. The active zone was placed at a specified co-localization distance, $l_c$ (center-to-center distance: 20 nm-400 nm) from the VDCC cluster (source of $Ca^{2+}$ flux) [8]. This canonical presynaptic terminal was implemented in a rectangular box 0.5 μm wide and 4 μm long representing a segment of axon making an en passant synapse. Calcium buffers modify the calcium diffusion rate and ultimately the local calcium profile. The diffusion length for calcium ions in our system was measured over several hundred trials and fit to the diffusion equation to calculate the effective diffusion constant. This was ~50 $\mu m^2/s$, close to experimentally measured values [15] (compared to the free diffusion constant of ~220 $\mu m^2/s$) and our local calcium profiles compare well with other studies [6](Fig. S2). The results on vesicular release rate presented here are therefore independent of the details of the buffering as long as the effective diffusion constant of calcium is maintained.

Exocytosis is primarily governed by the VDCC calcium currents. The arrival of an axonal action potential in the presynaptic terminal leads to the stochastic opening of VDCCs. . The total calcium flux entering the terminal depends on the time course of the action potential, the number of channels present on the membrane, the calcium conductance of open channels, and the total time each of the channels remains open. The calcium ions diffuse away from their point of entry into the terminal, where they may encounter and bind to calbindin, the calcium sensors and the PMCA pumps. A vesicle release takes place if sufficient calcium ions bind to the calcium sensor enabling the sensor to transition into an appropriate active state (Fig. 1B). The co-localization distance ($l_c$) and the calcium flux entering the presynaptic terminal tightly regulate the local calcium profile at the active zone and therefore control the neurotransmitter release probabilities. Every run was initiated with an action potential, which set the stage for the rest of the events leading up to success or failure of vesicular release depending on the specific $l_c$ and number of VDCCs (typically, a few tens of channels). Because the simulations were stochastic, we performed 10000 trials of each test case to generate an average release profile that could be compared directly to experimental data.

Release at a single active zone with seven docked vesicles was governed by a dual calcium sensor kinetic scheme (Fig. 1B). Simultaneous release of multiple vesicles was prevented by imposing a refractory period of 6 ms after a release event takes place [16,17]. The dual sensor kinetic scheme used in the simulations was similar to that proposed by Sun et al [18], in which one of the sensors regulates synchronous release via Synaptotagmin II (Syt II) and has 5 calcium binding sites, while the other regulates slow, asynchronous release via an as yet unidentified molecule and has 2 binding sites. Our attempts to reproduce the asynchronous release were most successful when 2 binding sites were implemented for the second slow sensor. At the calyx of Held, it has also been postulated that vesicles at active zones that are located farther from the calcium source [19] may release more slowly, leading to a longer time scale of release. However, for a CA3-CA1 synapse, which typically has only a single active zone, geometry alone cannot account for the asynchronous release. Moreover, experiments in which the fast sensor is knocked out continue to show asynchronous release transients, suggesting a model in which asynchronous release has a second slow sensor [18]. The similarities and differences between our model and that of Sun et al. are outlined as follows. In Sun et. al [18] the two sensors act completely independently to cause release and all releases are independent events. In contrast, in our kinetic model for CA3-CA1 the release of one vesicle (whether synchronously or asynchronously) temporarily prevents the release of other vesicles within the active zone. A refractory period results with a recovery time constant of ~6ms [16,17]. Our model differed from Sun et al. [18] in the binding and unbinding rates while maintaining the binding affinity and cooperativity of the calcium sensor for synchronous release. To better match published data [2] the asynchronous release in our model lasted much longer and had a much higher amplitude suggesting that this synapse has a longer memory. This was achieved in the model by making the unbinding rate of the second sensor 5 times slower than that in Sun et al [18]. Another significant distinguishing feature of the present model is that it includes a readily-releasable pool (RRP) with 7 docked vesicles [1], which is decremented after a release. This feature allows the model to accurately describe plasticity phenomenon such as depression and facilitation. All the results described below unless explicitly stated remain valid for a range of typical RRP sizes (results not shown). In our model, the vesicle fusion rate for asynchronous neurotransmitter release was not the same as the synchronous vesicle fusion rate ,$\gamma$, [6], as was reported in Sun et al. [18]. Identical fusion rates for both sensors leads to inconsistencies, as discussed in detail in the results section.

## Results

Figure 2A shows the neurotransmitter release probability as a function of the peak of the local calcium transient (measured at 10 nm from the sensor) for multiple co-localization distances ($l_c$). The number of VDCCs present in the cytoplasmic membrane regulated the calcium flux at the specified $l_c$. Small $l_c$ led to sharper, narrower local calcium peaks at the active zone (see Fig. S3) and the response curves for different $l_c$ were non-overlapping (Fig. 2A).

The base level of neurotransmitter release rate, in the absence of a stimulus, gave a true measure of the sensitivity of the calcium sensor that was independent of $l_c$ and other ultra-structural details. The spontaneous release rate in our model (Fig 2B) matched the release rate of $10^{-5}$ to $10^{-4}$ per ms reported in recordings from CA3-CA1 [20]. This agreement validates the values chosen for the forward and backward binding rates of the calcium sensor.

A response to a single action potential produced a 400 msec long elevated release rate of neurotransmitter at CA3-CA1 synapses [2] and exhibited two decay time scales as observed experimentally (see Fig.3A.): 5-10 ms ($\tau_{fast}$) and 100-200 ms ($\tau_{slow}$) [2]. Two time scales for decay (slow component of ~82 ms) have also been reported in parvalbumin-containing GABAergic interneurons expressing P/Q calcium channels [21].

The model correctly captured the release profile of hippocampal neurons reported by Goda and Stevens [2] shown in Fig. 3A. The response to an action potential averaged over 10,000 trials in 10 ms bins (Fig. 3C, black line) gave decay time constants of $\tau_{\mathrm{fast}}$ (7.4 ms) and $\tau_{\mathrm{slow}}$ (163 ms) [2], in agreement with the reported data (Fig. 3A). There was an increase in the overall contribution of asynchronous release ($\tau_{slow}$). An increase in the rate of decay of the synchronous release ($\tau_{fast}$) as well as decrease in asynchronous release ($\tau_{slow}$) compared with using the unmodified rates of the dual sensor model of Sun et al [18] in our spatially-extended synapse geometry (Fig. 3C, grey line). The affinity of the synchronous pathway, 38 μM, was the same as that in Sun et al. [18], which implies that both the calyx and the CA3-CA1 synapse have the same calcium sensitivity for release, since the fast component contributed more than 90% to the overall release probability (Table 1). Fig. 3D (red line) shows the local $[Ca^{2+}]_i$ 10 nm from the active zone (units on right-hand axis of graph). The neurotransmitter release peaked after a typical latency of ~ 3 ms. Note that here we measured the latency starting from the beginning of the action potential (i.e. t=0 in Fig. 3D is at the beginning of the action potential), This latency is due mainly to the delay in opening the VDCCs as the action potential depolarized the axon. The local $[Ca^{2+}]_i$ peaked at ~11 μM for $p_r$ =0.2.

When the data were binned at 1ms (Fig. 3D black line, units on left-hand axis), a third super-fast timescale of release was apparent that had a time constant of $\tau_{superfast}$ = 0.526 ms and was highly correlated with the time course of the $Ca^{2+}$ pulse (Fig. 3D red line, units on right-hand axis). This follows the vesicle fusion rate $\gamma$ that is fast enough to track the calcium transient. A rate of 2000/s was sufficient to track the calcium profile of fast P/Q calcium channels in our model, and was less than the measured release rate of 6000/s [6,18]. This super-fast timescale of release has been observed in calyx of Held (Fig 3B) by Scheuss et al. [22]. Their 'biphasic decay of release rate' was comprised of a superfast component of release and a fast component (588.6 ±3.5 μs and 14.7±0.4 ms respectively). However, they were unable to distinguish the contribution of slow asynchronous release lasting up to 200 ms, from the effect of residual glutamate in the cleft. Thus, several different times scales of release by different labs ($\tau_{fast}$ and $\tau_{slow,}$) or ($\tau_{superfast}$ and $\tau_{fast}$) have been reported [2,18,21,22]. This disagreement can be explained by the coexistence of three time scales of release in the CA3-CA1 synapse, as seen in Figs. 3C and 3D.

Our prototype CA3-CA1 synapse achieved $p_r$ = 0.20 with 48 VDCCs in a single cluster of 35 nm radius, at $l_c$ = 250 nm, which is compatible with estimates made at other central synapses [8]. This was not a unique model since other combinations of VDCC number and $l_c$ also gave $p_r$ = 0.20, without any significant modification of our findings. Most hippocampal synapses have a low probability of release and have an average baseline value of $p_r$ ~0.2 [3]. One estimate of the average release probability per active zone at the calyx was also ~ 0.2 [23], suggesting similarities between the two synapses. However, the range of release probabilities at hippocampal synapses is high, from weak synapses ($p_r$ < 0.05) that rarely ever release to synapses with high release rates ($p_r$ > 0.9) [3]. As illustrated in Fig. 2A, release probability is a function of local calcium concentration at the active zone and can be modulated by either varying the number of VDCCs or the distance between the calcium source (VDCC) and the calcium sensor ($l_c$ ) in our simulations. RRP size is an additional way to modulate release probability but recall that we fixed the initial RRP size at 7 vesicles. We find that the maximum amplitudes of the synchronous and asynchronous contributions were modulated by the varying $p_r$, but the decay constants of the release profiles were unchanged (Fig. 3E; $p_r = 0.6$, $l_c$ =400 nm, 128 channels; Fig. 3F; $p_r = 0.92$, $l_c$ =250 nm, 112 channels). This result of the model is consistent with reported data from high and low release probability synapses that show similar decay constants [2,22] for different release probabilities.

For the calyx of Held synapse, the slower time scale of release has been attributed to active zones farther from the calcium source ($l_c$), compared with faster release from vesicles located close to point of calcium entry [19]. In our simulations, the decay time scales were independent of the spatial organization of the synapse and were a consequence of the kinetics of the calcium sensor (See Fig. 3E). A recent study reports independence of specific properties of the $Ca^{2+}$ channels and relative location of $Ca^{2+}$ in shaping the relative dynamics of asynchrony to phasic release, further corroborates our result [24]. This result of the model also supports an after-release refractory period, to be discussed in detail later.

The independent contributions of synchronous and asynchronous release are shown in Figs. 4A-C. Initially, the fast (and superfast) release dominates, but it decays rapidly and is soon overtaken by asynchronous release. The synchronous part of the release is the primary contributor to the $\tau_{superfast}$ time scale, which is called ‘phasic synchronous release’; the $\tau_{fast}$ time scale is also mainly driven by the synchronous pathway and is called ‘delayed synchronous release’; finally, the $\tau_{slow}$ release is called ‘asynchronous release’. The asynchronous contribution to the release profile has a delayed peak compared to the synchronous contribution, which is also present in the data from Sun et al. and Otsu et al.[18,25].

As mentioned above multiple releases can take place from the active zone after a refractory time constant of ~ 6 ms following each release [16,17]. To test if the finite available resource of docked vesicles (i.e. the RRP) is a limitation, we simulated an active zone in which a released vesicle was instantly replaced, i.e. a depletion free active zone. The probability distribution of number of quanta of neurotransmitter released in 400 ms is shown in Figs. 4D-F. For a synapse with a release probability, $p_r = 0.2$ (48 VDCCs at 250 nm) the likelihood that more than two vesicles were released was less than 5% (from cumulative release probability plot. Furthermore, there was less than 20% chance of releasing more than 2 and almost never more than 6 vesicles for $p_r = 0.6$ and a 33% chance of releasing more than 2, and almost never more than 9 vesicles for $p_r = 0.95$. The size of readily release pool (RRP) has been estimated to be 5-10 vesicles at CA3-CA1 synapses [1]. That the maximum number of vesicles released is consistent with the typical RRP size at this synapse and that both these numbers are positively correlated with release probability [26] is additional support for our modeling framework. The model further suggests that the typical RRP size at a CA3-CA1 synapse and the calcium sensitivity of the release machinery are well-matched, so that the number of docked vesicles is not a limiting factor at low stimulus frequencies.

## Refractory Period

Stevens and collaborators introduced the idea that there is a short refractory time constant following vesicle release from an active zone. With such a refractory period more than one quantum of neurotransmitter can be released by an action potential, but the quanta are released one at a time. Several recent experimental studies have tried to address the question of refractoriness after release but with conflicting results. Explicit measurements at a wide variety of synapses conclude that there exists a “one active zone-one vesicle release” principle and hence provide direct evidence for functional coupling within the active zone [16,17,27,28,29,30,31,32,33,34,35]. However, other studies have presented evidence against uni-vesicular release due to such “lateral inhibition” [28,36,37,38,39,40,41,42,43]. Our basic strategy is to compare neurotransmitter release profiles with and without the existence of a 6 ms refractory time constant preventing simultaneous release of different vesicles. We do this for different values of the overall release probability (See Fig. 5).

For a release probability at CA3-CA1 of $p_r$=0.2, the release transient for a synapse with a refractory period (gray line) is almost indistinguishable from a synapse without any refractoriness (black line). Thus for this set of parameters, the presence or absence of refractoriness does not make any functional difference. For a release probability of $p_r$ = 0.2 for the whole active zone, each of the 7 individual docked vesicles must have a release probability of 0.035 so the probability that 2 or more vesicles being released is only 0.02. This implies that although any single vesicle was released on 20% of the stimuli, two or more vesicles were released on only 2% of the trials. The detailed timing of release of the second vesicle relative to the refractory period has a negligible effect on the overall averaged release profile. The consequence of a refractory period was more prominent for $p_r$ = 0.95. The high release probability was implemented by increasing the number of VDCCs to 112 at $l_c$ = 250 nm, with all other parameters, including the calcium sensor, exactly the same. Now, for a synapse with independent releases (i.e. no refractory period) and $p_r$ = 0.95, 2 or more vesicles were released on 67% of the trials. The top panel in Fig. 5B shows the release transients over 400 ms when the release data were in 10 ms bins and the bottom panel (Fig. 5D) describes the same data with finer 1 ms bin. Now, there is a clear consequence to the inclusion of a refractory period.

An important distinguishing characteristic of neurotransmitter release in hippocampal CA3-CA1 synapses, the calyx of Held, and parvalbumin interneurons is that the two decay time scales are conserved across a wide range of release probabilities even as the overall amplitude of the transient is modulated [21,22], This observation could be replicated in our model only when refractoriness was included. Without refractoriness, depletion overwhelmed the release at high release probability synapses: The peak release rate was higher, the decay was faster and the amplitude of later releases was lower (Fig.3F, black line).

We next examined the differences in the release transients due to refractoriness separately for the synchronous and asynchronous release for $p_r$ = 0.95 (see Fig.6A and B). This analysis was possible because our sensor model treated these releases via independent pathways reaching activated states (see Fig 1B). Our model predicted that the synchronous release profile (Fig. 6A) should be lower in amplitude and decay more slowly for a synapse with a refractory period. Synchronous and asynchronous releases compete for the same RRP resources [25] leading to a net increase in asynchronous release (1511 total events in 400 ms) for the synapse with refractoriness compared to the synapse without refractoriness (1379 total events in 400 ms) (Fig. B). Note that in the first ~50 ms after the stimulus, when release via the synchronous pathway dominates, refractoriness slows the rate of depletion of the RRP (Fig. 6A) and thus allows synchronous release to initially suppress asynchronous release (Fig. 6B). But beyond 50 ms, when asynchronous release begins to dominate, the larger residual RRP in synapses with refractoriness means that the net amount of release via the asynchronous pathway can be larger than in synapses without refractoriness.

Gene knock-out experiments are now routinely used to quantify signalling pathways. Knocking out synaptotagmin II, the calcium sensor for neurotransmitter release, eliminated the fast release component of the transient but left the slow component intact [18,44]. Augmentation of asynchronous release in genetically modified, fast sensor deficient mice has been previously reported [44]. We modified our model to allow for the study of the KO transgenics by removing all the states along the synchronous pathway. Since both pathways used the same resource pool of neurotransmitter [25], knocking out the synchronous release sensor made more vesicles available for release through the asynchronous release sensor. Simulation results for asynchronous release transients comparing synchronous sensor knock-out (KO) and wild type are shown in Fig. 6C and D. The results show that the genetic modification eliminates much of the effect of the refractory period (gray solid line and black solid line respectively) with almost the same number of release events for both in the 400 ms (inset) and 50 ms time windows. The genetic modification had a larger effect on the refractory synapse and was qualitatively more consistent with experimental data.

For a synapse without refractoriness the difference (Fig. 6D) between the release rate of the wild type and KO stayed constant through the transient; however, for a synapse with refractoriness (Fig. 6C), the model predicted that the difference between wild type and KO would be larger in the first few milliseconds and taper off with time. This was because the large forward binding rate of the synchronous part of the sensor dominated release in the wild type and therefore acted to inhibit asynchronous release; this inhibition occurred through refractoriness that lasted a few milliseconds before the asynchronous channel could reach its normal release rate as defined by the binding kinetics. The increase in release rate of asynchronous release in first 50 ms reached 90% in a synapse with refractoriness compared to an increase of 75% for the synapse without refractoriness. In a synapse without refractoriness, synchronous and asynchronous releases were independent and therefore they would always occur at their normal rates.

Refractoriness differentially affects synchronous and asynchronous release at early and late times after a single stimulus and this effect is sensitive to the initial release probability (Fig. 5). But what happens during a train of high-frequency stimuli? We performed simulations to predict what might be seen in CA3-CA1 synapses when stimulated at 100 Hz for 200 ms (20 stimuli) and examined the results for features that would distinguish between synapses with and without refractoriness. This same stimulus protocol was used in a previous study of the calyx of Held [22] and was found to be sufficient to deplete the RRP. We surmised that such a stimulus might also be sufficient to deplete the RRP at our model CA3-CA1 synapse with a single active zone – if it is reasonable to think of the calyx of Held as a few hundred active zones, each containing ~7 docked vesicles responding independently to the same stimulus.

The response of our model synapse with initial release probabilities of $p_r$ = 0.2 (number of VDCCs =48, $l_c$ = 250 nm), $p_r$ = 0.6 (number of VDCCs =72, $l_c$ = 250 nm), and $p_r$ = 0.9 5 (number of VDCCs =112, $l_c$ = 250 nm) is shown in Fig. 7. For $p_r$ = 0.6 the facilitation (ratio of first two release rates) in the synapse with refractoriness (black line) was almost twice that of a synapse without refractoriness (grey line). However for the synapse with refractoriness the background release level (due to asynchronous release) was much higher compared to a synapse without refractoriness. These results can be directly tested in hippocampal synapse experiments.

**Sensor memory and short term plasticity**

Response to 10 Hz train stimuli for 400 ms for a synapse with intrinsic release probability 0.2 is shown in Fig. 8. The simulations are carried out for an asynchronous sensor KO (Fig. 8B) and wild type (Fig. 8A). The response to higher frequency (100 Hz) is discussed in the supplementary material. Unlike the KO (Fig. 8B), the peak release rate (data binned in 1ms) in the wild type (Fig. 8A) is facilitated with each subsequent stimulus. The same data (grey line- asynchronous sensor KO, black line- wild type) is shown on a log scale in Fig. 8C. In the wild type, response to subsequent stimuli, ride on top of the higher base level release. This is due to the slow time scale of release of the asynchronous sensor (the inherent memory of the sensor). This ensures greater facilitation for the wild type. Fig. 8D shows the total release rate for each stimulus (grey line-asynchronous KO and black line – wild type). We can see that for the facilitation in the wild type is more than 50% whereas for the asynchronous KO it is limited to 35%.

**Vesicle Fusion Rates**

A sample release profile of the asynchronous pathway for our single active zone synapse with 7 docked vesicles [1] assuming equal release rates for both release pathways is shown in Fig. 9 (circular glyphs, $p_r = 0.2$, number of VDCC = 48, $l_c = 250$ nm). The early peak in this figure, for simulations at all values of the release probability, was clearly inconsistent with electrophysiological data [18,25]. We were unable to eliminate this early peak in the asynchronous release while still reproducing all the other measured release properties either by changing the binding affinities or by including additional calcium binding sites for the asynchronous pathway that would delay release (data not shown). In order for our model to be consistent with measured asynchronous release transients, the value of $\gamma$ needed to be 40 times slower for the asynchronous pathway relative to the synchronous pathway. This introduces an additional parameter '*a*' such that the neurotransmitter fusion rate is '$a\gamma$'.(with $a < 1$) for asynchronous release (see Table 1). The presence or absence of refractoriness did not affect this early peak through the asynchronous pathway. For $a$ = 0.025 (i.e. net asynchronous vesicle fusion rate = 50/s), the early phasic release from the asynchronous pathway was suppressed and all the detailed characteristics of neurotransmitter release were reproduced (Fig. 9, square glyphs).

An alternative way to eliminate the early peak in the asynchronous release while implementing neurotransmitter fusion rates for synchronous and asynchronous release was to use a phenomenological model for the entire active zone such that it has a single gating mechanism prescribed by kinetic rates given in Table. 1. This model sets no *a priori* limit on the number of docked vesicles (i.e. has an infinite RRP) and multiple release events may occur subject to the refractory time constant.. With this model it was possible to reproduce all data consistently including the 3 time scales and cumulative release well matched to the RRP (data not shown). In summary, an additional parameter '*a*' was needed in the docked vesicle model with individual sensors on each vesicle, to directly suppress asynchronous release, whereas in the phenomenological model that treats the whole active zone as having a single gating mechanism, no such parameter was needed.

## Discussion

Neurotransmitter release at chemical synapses in response to a stimulus is tightly regulated over multiple time scales by mechanisms in the presynaptic terminal. Release takes place at specialized locations at the presynaptic membrane called active zones designated by the presence of SM (Sec1/Munc18-like) proteins [45,46]. Some of this machinery is ubiquitous for all exocytosis events and consists of SNARE (soluble N-ethylmaleimade-sensitive factor attachment protein receptor) proteins, SM (Sec1/Munc18-like) proteins, along with complexins and synaptotagmins that are needed to control the timing of neurotransmitter release [45]. Much of the molecular and structural details of this process have been elucidated; however, how each of the components interacts to execute precise dynamic control on the release has not yet been established. The goal of this study was to begin developing a detailed biophysical model of exocytosis that takes into account the spatial organization of the molecular components and the time courses of their kinetic states.

**Synaptic transmission at small synapses**

Vesicular release at synapses has been studied in great detail over the last few decades to understand the cellular basis of plasticity and higher brain function. These studies have not always been in agreement, which has led to confusion about certain essential aspects of synaptic transmission. Our computational experiments have led to possible resolutions for some of these contentious issues, such as the existence of refractoriness between releases, cohesively bring together data from different sources that point to universal features of vesicle release and those that may be unique to CA3-CA1 synapse [47,48]?

In particular, our simulations have illuminated the observation in two separate sets of data [21,22] that changing the release probability modifies only the amplitudes of release transients and not the timing of release. An important prediction of this study is that the three decay time scales of release are independent of the synaptic geometry. It has been reported in a recent study [24] that properties of the $Ca^{2+}$ channels and relative location of $Ca^{2+}$ do not modulate the relative dynamics of asynchrony to phasic release, suggesting a differential mechanism for both. This study strongly supports our own modelling results.

Synaptotagmin I/II are ubiquitous calcium sensor proteins at synapses that most likely govern fast release in central synapses, but the component that governs asynchronous release is still unknown [18,49]. In our model of the calcium sensor, the synchronous part of the machinery that determines the probability for fast release, has exactly the same affinity (38 μM) as a previous model of the calyx put forth by Sun et al. [18]. The $Ca^{2+}$ sensitivity for each active zone in the calyx is comparable to that at hippocampal synapses [23], but the CA3-CA1 synapses show a slow component of release in the model that lasts much longer and has higher amplitude, implying a longer sensor (Fig. 3C) memory.

The calyx and the CA3-CA1 synapses subserve different functions. The calyx is a giant synapse in the auditory pathway that achieves reliable synaptic transmission with several hundred active zones. In comparison, most CA3-CA1 synapses in the hippocampus have an intrinsically low release probability but are highly plastic [26] to serve as a substrate for memory [50,51]. Despite these differences, the calcium sensor that governs fast temporally correlated signal transmission seems to be conserved. Asynchronous release transients may be more diverse, although at a particular calyx synapse that exhibited an exceptionally high level of asynchronous release, Scheuss et al. [22] reported a slow asynchronous decay with a time scale that was comparable to that in our model (79.3 ±29.7 ms). Furthermore, the global parameters of the synapse, such as the number of active zones, and their respective distance from the VDCCs, can give rise to apparently different calcium sensitivities that can be misleading (see Fig. 2B). Whether universal or not, a $Ca^{2+}$ sensor with a long memory as described in our hippocampal model can have a significant role in activity-dependent short-term synaptic plasticity (Fig. 8).

**Refractoriness**

The active zone is morphologically distinctive and has complex protein meshes spanning the entire length of the region connecting all the vesicles [52]. Recently, a diffusive protein trans-complex was identified that forms a continuous channel lining at the fusion site and is integral to exocytosis [53]. A local perturbation caused by exocytosis is likely to be spread through these diffusive molecules. It has also been suggested that the mechanical rearrangement of the lipid bilayer during exocytosis can also affect later release over a short enough time scale [54]. Given all these opportunities to influence each other, there are likely to be conditions under which docked vesicles interact cooperatively.

Our simulations suggest that the release of a vesicle may trigger direct and indirect exhanges between the synchronous and asynchronous release pathways, between individual sensors on the several docked vesicles, and between the microenvironment of the membrane of the active zone and the vesicles. These interactions occur on several time scales. In the model, "Lateral inhibition" a refractory period with a time constant of 5-7 ms [16,17,55] blocks simultaneous release from the active zone during the period of highest calcium concentration after opening of VDCCs. The exact biophysical mechanism for this refractory time window is unknown.

The times scales of decay measured in the release transient is are conserved across release probabilities and only the amplitude of the response is modulated with the probability of release [21,22]. Without a refractory period of 6 ms after a release event, it would not be possible to maintain the same decay time scales across all release probabilities (compare $p_r$ = 0.2 and $p_r$ = 0.9 shown in Fig. 5). In addition, the prediction of the facilitation and base level release as illustrated in Fig. 7 can also be rigorously tested experimentally for further confirmation and exploration of the phenomenon..

Some of the discrepancies leading to different conclusions about the refractoriness following vesicle release [16,17,27,28,29,30,31,32,33,35,36,37,38,39,40,41,43,56] could be due to differences in techniques and stimulation protocols. The proposed refractoriness originally measured by Dobrunz et al.[16] lasted only a few ms and did not impede subsequent release beyond that time window. Oertner et al. reported multivesicular release accompanied by an increase of glutamate in the synaptic cleft. It is possible that more than one vesicle was indeed released but separated in time by the refractory period, since their methods lacked temporal resolution to resolve millisecond differences.

Simultaneous release within synapses containing more than one active zone is also possible [36,39]. We have estimated that if release indeed operated independently at each docked vesicle, for $p_r$ = 0.9 there should be a 70% chance of releasing more than 2 vesicles in response to a single action potential, but in Christie et al [37] multivesicular release was observe only in a paired pulse facilitation protocol.

The accumulation of glutamate in the synaptic cleft could also give a misleading interpretation of multivesicular release. Abenavoli et al [38] performed statistical analysis of minis where they observed that the output at long time intervals was not Poisson distributed. This phenomenon was attributed to burst of release from the same synapse that contradicts the refractoriness and led them to conclude that multivesicular release occurred at the CA3-CA1 synapse. An alternative explanation is the existence of long-time correlations in neural activation, perhaps by astrocytes acting to synchronize activity [57,58]. Furthermore, the quick freeze technique they used to image synaptic vesicles did not have the temporal resolution to distinguish between endocytotic and exocytotic events.

Asynchronous release is enhanced in transgenic mice with the fast sensor knocked out at synapses [44]. In addition to making more vesicles available for the asynchronous pathway, another recently proposed mechanism is that of zipping action of complexins with synaptotagmins that clamps down release in the wild type [59]. Binding of calcium releases the complexin clamp. However, in the synaptotagmin KO this clamp is abolished, leading to an increase in asynchronous release. Further experiments will be needed to test whether this more detailed mechanism occurs, given that we can already obtain significant augmentation from the existing model.

Our approach to exploring the contentious issue of whether or docked vesicle are independently released has been to model a single active zone and make predictions for the different possibilities that can be compared with published data and tested in further experiments. Since these CA3-CA1 hippocampal synapses typically have only a single active zone, we have dissected the contributions made by synchronous and asynchronous release and have been able to show how release is modulated by a refractory period. Many of the properties that have been observed in nerve terminals can be explained in our model by assuming a refractory period for vesicle release, which implies some form of coupling between docked vesicles at the active zone. Further work is needed both to test some of the model predictions regarding facilitation and to determine how the vesicle coupling arises via molecular mechanisms.

**Neurotransmitter Fusion Rate**

The model has an active zone with a RRP of vesicles that are coupled through a brief refractory period following each release via either the synchronous or asynchronous pathway. This differed from kinetic models for the calyx of Held [6,7], including that of Sun et al.[18], which assumed that every vesicle release was independent. In the calyx, Sun et al. used the same vesicle fusion rate ($\gamma = 6000\ s^{-1}$, see kinetic scheme in Fig.1) as measured by Schneggenburger and Neher [6] and made this rate equal for both the synchronous and asynchronous pathways. The slow-to-release vesicles showed the same release transients as seen by Wadel et al [19] when calcium was un-caged so that calcium concentration was uniform across the presynaptic terminal of the calyx. This suggests equal neurotransmitter fusion release rates, $\gamma$, since in calcium-uncaging protocols, it is likely that calcium ion binding is not the rate-limiting quantity.

However, it was only possible to fit all the release data for CA3-CA1 synapses when we set the value of the neurotransmitter fusion rate, $\gamma$, to be 40 times slower for the asynchronous pathway relative to the synchronous pathway. An alternative possibility is that there might be additional coupling in the active zone beyond the refractoriness that makes the active zone behave as if there were a single gate. The overall effect of this inhibitory coupling is to reduce the effective asynchronous neurotransmitter fusion rate, which should be pathway-independent. Developing this possibility further would require a better understanding of the proteins that are responsible for the coupling and the concomitant explicit sensor-sensor coupling in the kinetic scheme. Experimentally, one would need to develop knock-outs of the coupling proteins and test these for evidence of enhanced asynchronous release rates, especially the existence of an early release peak not present in wild-type synapses.

## Materials and Methods

Simulations were performed using MCell, version 3 [60,61]. MCell uses Monte Carlo algorithms to simulate volume and surface reaction-diffusion of discrete molecules in complex spatial environments with realistic cellular and sub-cellular geometry. This allows for detailed study of the effect of the spatial organization and stochastic reaction-diffusion dynamics on the temporal evolution of key system variables. We modelled a 0.5 μm × 0.5 μm × 4 μm volume of simplified en passant axon segment with physiologic spatial distributions and concentrations of ligands and molecules. Initial concentrations, locations, diffusions constants, and rates and their sources used for the MCell model are specified in Table 1. Further validation of the parameters used comes from the shape and amplitude of the calcium response to action potential in our simulations which is consistent with experimental data [10,11].

The apparent diffusion constant of calcium, a key parameter for physiological relevance of our results, was matched in the model to the measured value (50 $\mu m^2$/Sec) [15]. This value is substantially slower than the initial free diffusion constant of 220 $\mu m^2$/sec specified for the simulation and arises because our model has an accurate description of the calcium binding kinetics of mobile calcium binding proteins in the synaptic. The calcium concentration was clamped at 100 nm at both ends of the axon segment. The simulation time step for calcium was specified to be 0.1 μsec and for all other molecules was 1.0 μsec. The release transients presented in the figures is a result of 10000 simulations for each parameter set. The docked vesicles were clustered in a hexagonal array with largest center-to-center distance between vesicles of 35 nm.

## Acknowledgements

We would like to thank Charles Stevens, Elaine Zhang, Dan Keller, and Donald Spencer for invaluable discussions and Jed Wing and Rex Kerr for software development. This work was partially supported by the Center for Theoretical Biological Physics (NSF PHY-

0822283), NIH (MH079076, GM068630, P01-NS044306), and the Howard Hughes Medical Institute.

Correspondence and requests for materials should be addressed to Terry Sejnowski. (e-mail: terry@salk.edu).

# Figures and Figure Legends

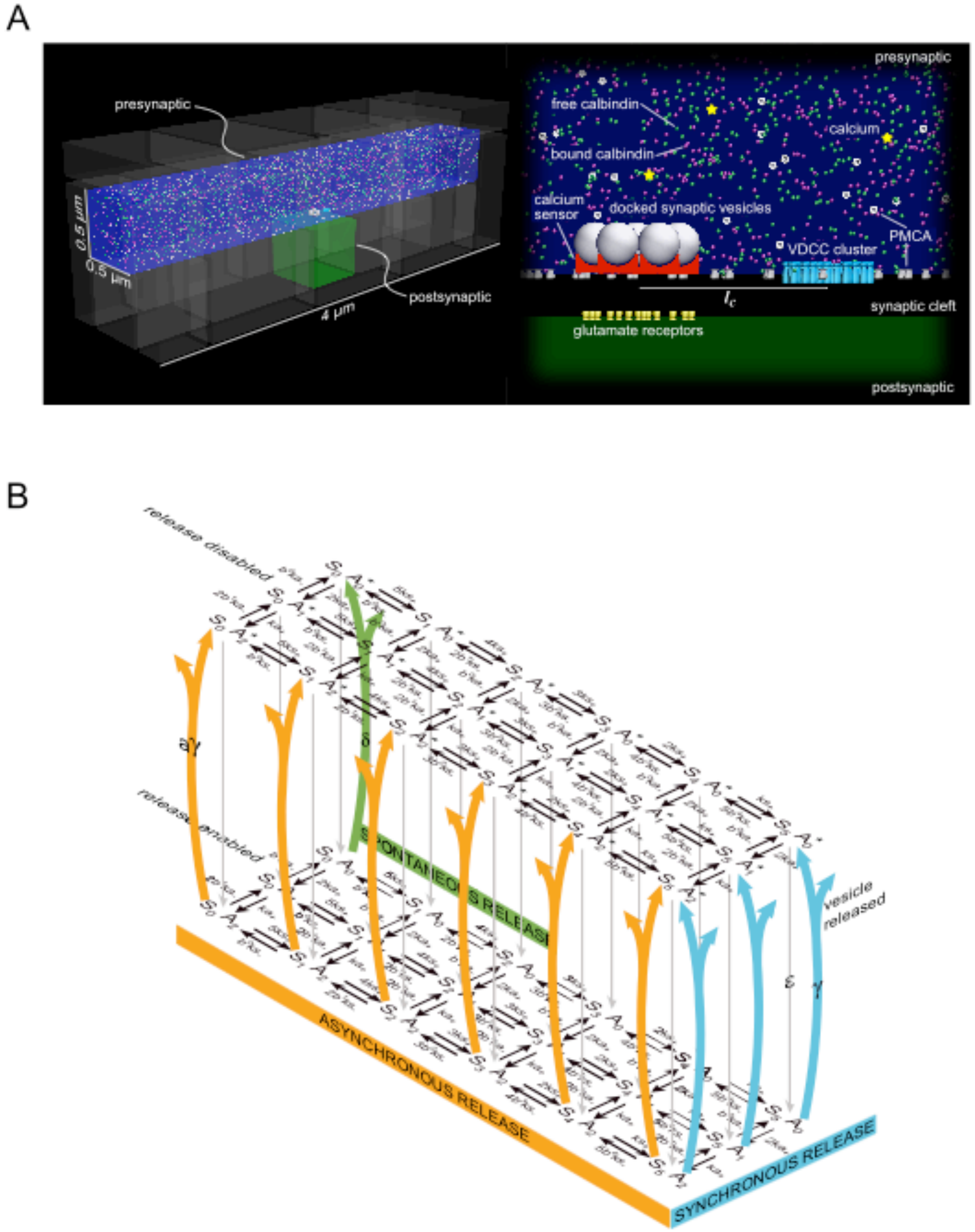

**Fig. 1 Canonical CA3-CA1 synapse. (A)** The model Shaffer collateral axon (blue) from CA3 making an en passant bouton (green) with the dendrite of a CA1 pyramidal neuron showing (right) the physiological spatial distributions and concentrations of ligands and molecules. The simulations were carried out in 0.5 μm × 0.5 μm × 4 μm volume of the axon including of a cluster of voltage dependent calcium channels (VDCCs), mobile calcium buffer calbindin and plasma membrane calcium ATPase (PMCA) pumps. The active zone was populated by seven docked vesicles each with its own calcium sensor for neurotransmitter release at a prescribed distance, $l_c$ from the VDCC cluster. **(B)** Kinetic model for the calcium sensor with 2 pathways, synchronous and asynchronous. The synchronous release pathway has five calcium binding sites whereas asynchronous release has two calcium binding sites. Note that the neurotransmitter release process has distinct rates, $\gamma$, for synchronous release and a slower one, $a\gamma$, for asynchronous release. When the refractory period was implemented, the release machinery was disabled after a release event takes place, whether via either synchronous or asynchronous, and was re-enabled with a time constant, $\varepsilon$, of 6.34 ms.

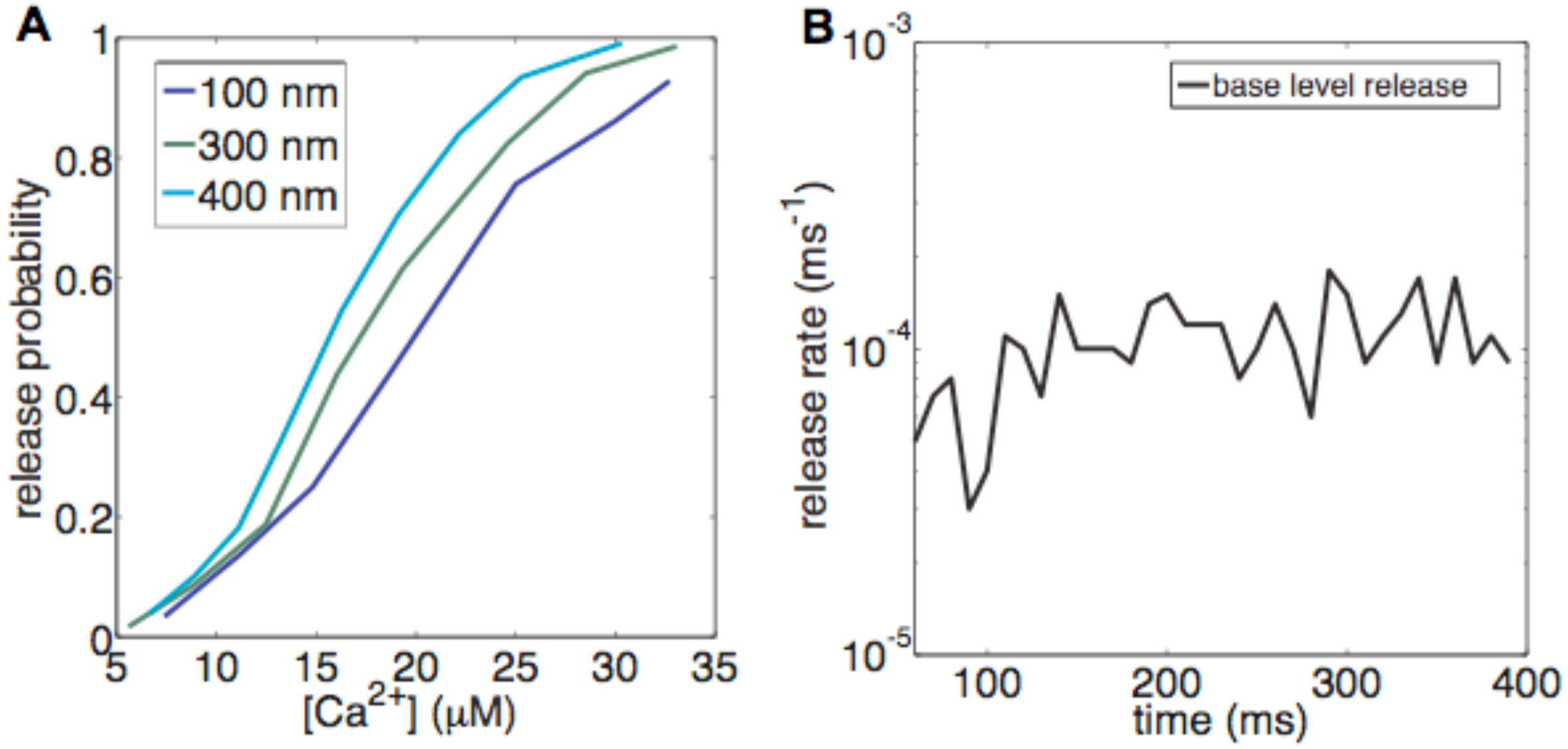

**Fig. 2.** **(A)**. Calcium sensitivity of neurotransmitter release response for a range of distances, $l_c$ between the calcium sensor and the VDCCs. A set of non-overlapping curves emerge for various distances. Local peak calcium concentration at the site of the active zone is a measure that is modulated by spatial details **(B)** The neurotransmitter release profile with no external stimulus illustrating the basal release rate. This steady state release profile is a distinct characteristic of the calcium sensor and is independent of geometry.

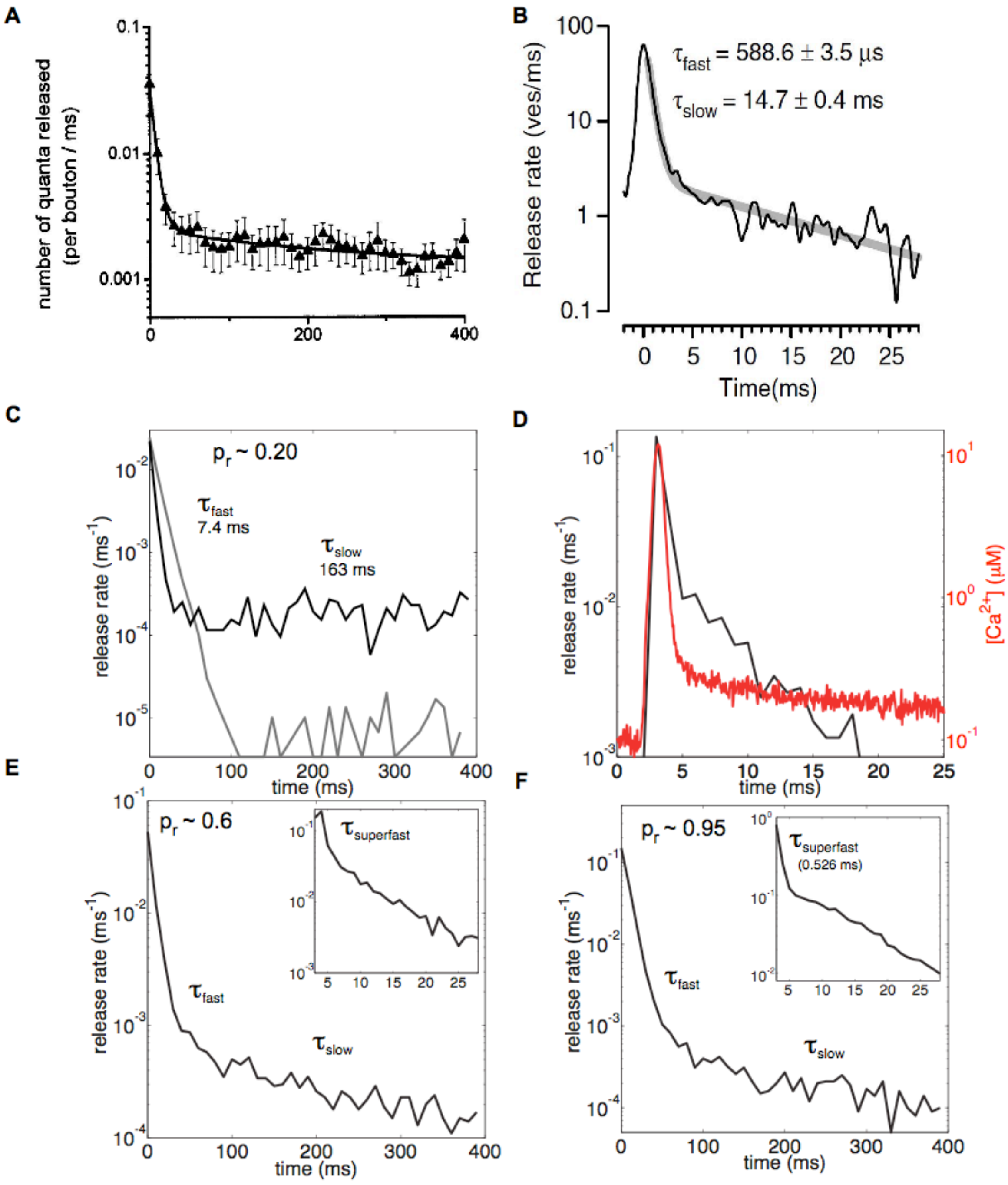

**Fig. 3 Quantal release time courses**. **(A)** Stimulus evoked neurotransmitter release data (Goda and Stevens [2] ) from dual patch clamp recordings in paired cells using hippocampal pyramidal neurons showing two time scales of release. **(B)** Data from Scheuss et al. [22] for measured release transient at the calyx of Held. **(C)** Black line

shows simulation of neurotransmitter release transient for a synapse with intrinsic $p_r = 0.2$ showing two distinct time scales of release (10 ms bins, compare with 3a). Grey line with shows simulations of kinetic model by Sun et al. [18] in a CA3-CA1 with a single active zone. **(D)**. A superfast time scale ($\tau_{superfast}$) emerges for neurotransmitter release rate ($p_r = 0.2$) using finer 1 ms bins (left axis, black line). Compare with the superfast timescale of release described at the calyx in 3b. The calcium pulse measured 10 nm from the calcium sensor in response to 48 VDCCs at $l_c = 250$ nm that triggered neurotransmitter release is superimposed (right axis, red line). The initial superfast part of the release is highly correlated to the calcium pulse (phasic synchronous release) and is followed by a fast timescale of release (delayed synchronous release). **(E,F).** Release transient in response to an action potential for synapses with $p_r = 0.6$ and $p_r = 0.95$ in 10 ms bins. The insets show the superfast timescale for the same data (1 ms bins). The release transient for $p_r = 0.6$ is generated for synapse with 128 VDCCs placed 400 nm from the sensor and 112 VDCCs placed at 250 nm for $p_r = 0.95$. Even though the maximum amplitudes of the two components of release in a $p_r$-dependent way, the 3 decay time constants $\tau_{superfast}$, $\tau_{fast}$ and $\tau_{slow}$ are conserved across a wide range of release probabilities.The decay time scales are also independent ultrasynaptic structure (compare b, d, e, f). For a synapse with $p_r = 0.2$ , 44% of release takes place at $\tau_{superfast}$, 43% at $\tau_{fast}$, and the remainder at $\tau_{slow}$. For comparison to Goda and Stevens (20) exponential decay times scales are fit to the equation $a_0 \exp(-t/\tau_{fast}) + a_1 \exp(-t/\tau_{slow})$ where $a_0 = 0.034$, $\tau_{fast} = 7.4$ ms, $a_1 = 0.00021$, $\tau_{slow} = 163$ ms. The 'superfast' timescale with 1 ms binning was fit to the equation $a_2 \exp(-t/\tau_{superfast})$ where $a_2 = 0.257$ and $\tau_{superfast} = 0.526$ ms.

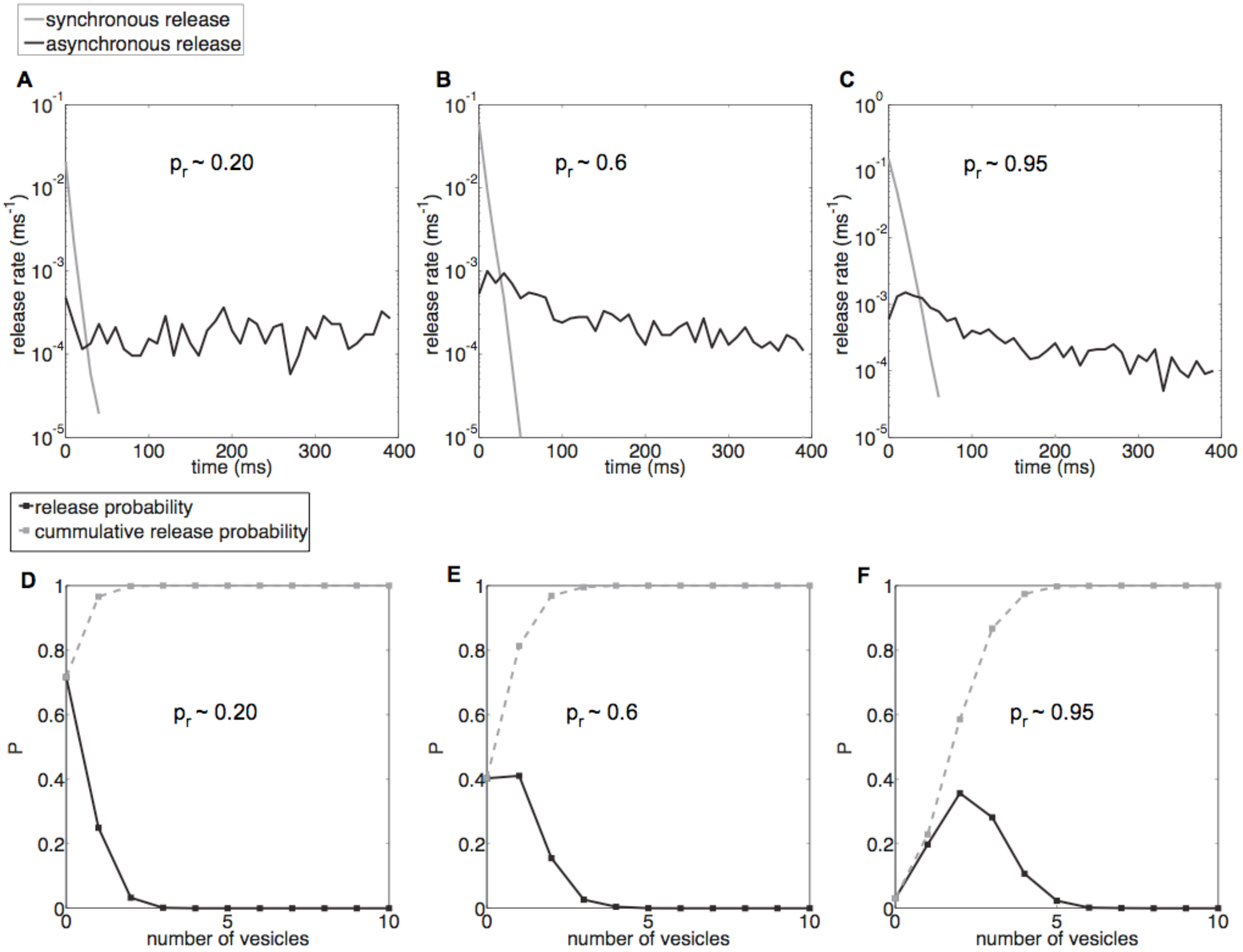

**Fig. 4 a-c.** Contributions of synchronous and asynchronous release for a range of probabilities. **(A-C):** The synchronous pathway is the main contributor of the phasic synchronous and delayed synchronous release. The asynchronous release peaks much later. **(D-F):** The probability distribution (black line) for the number of released vesicles when the RRP is set to be infinite (no depletion after release). Cumulative probability is shown in grey. Consistent with size of the RRP of CA3-CA1, more than 8 vesicles are rarely released. This validates the binding and unbinding rates of calcium ions for the sensor for vesicle release. Also synapses with higher intrinsic $p_r$ are more likely to release more vesicles per stimulus.

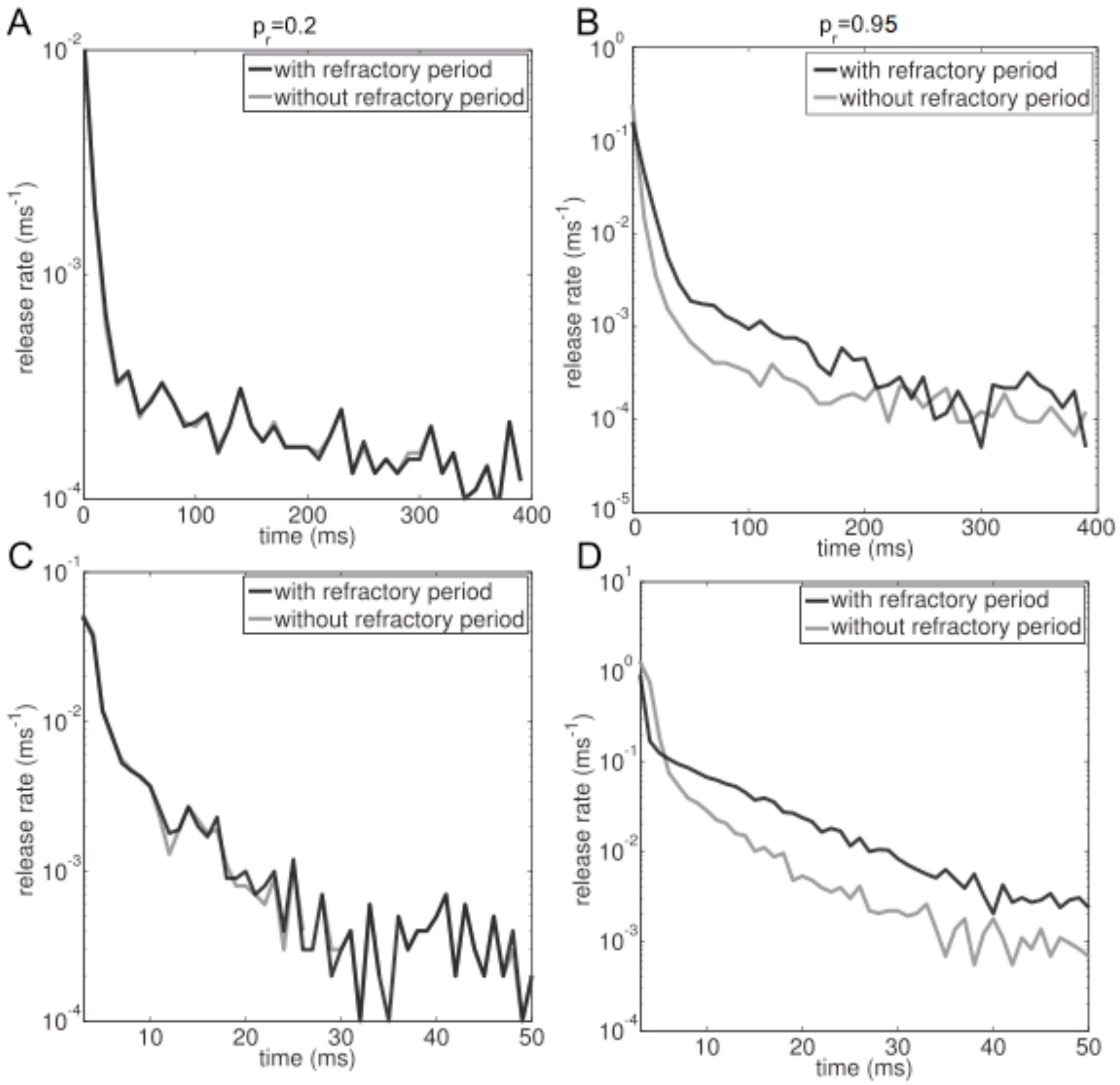

**Fig. 5 Neurotransmitter release profile for a CA3-CA1 synapse with a single active zone and seven docked vesicles. (A)** Release data histogram in 10 ms bins for a synapse with intrinsic release probability of $p_r$ = 0.2 (48 channels at $l_c$ = 250 nm). Both transient, refractory period transient (grey) and non-refractory period transient (black) almost exactly overlap. **(C)** This holds true for a finer 1 ms bin (bottom panel) as well. **(B)** Release data histogram in 10 ms bins for a high release probability $p_r$ = 0.92 (48 channels at $l_c$ = 250 nm). The two transients in this case decay with different rates. The synapse without the refractory period decays faster, as depletion of neurotransmitter vesicles cause decreasing release probability. **(D)** This effect is seen in more detail with 1ms bins at the same

synapse. Only for the synapse with refractory period are the characteristics time scales of decay conserved across the whole range of release probability.

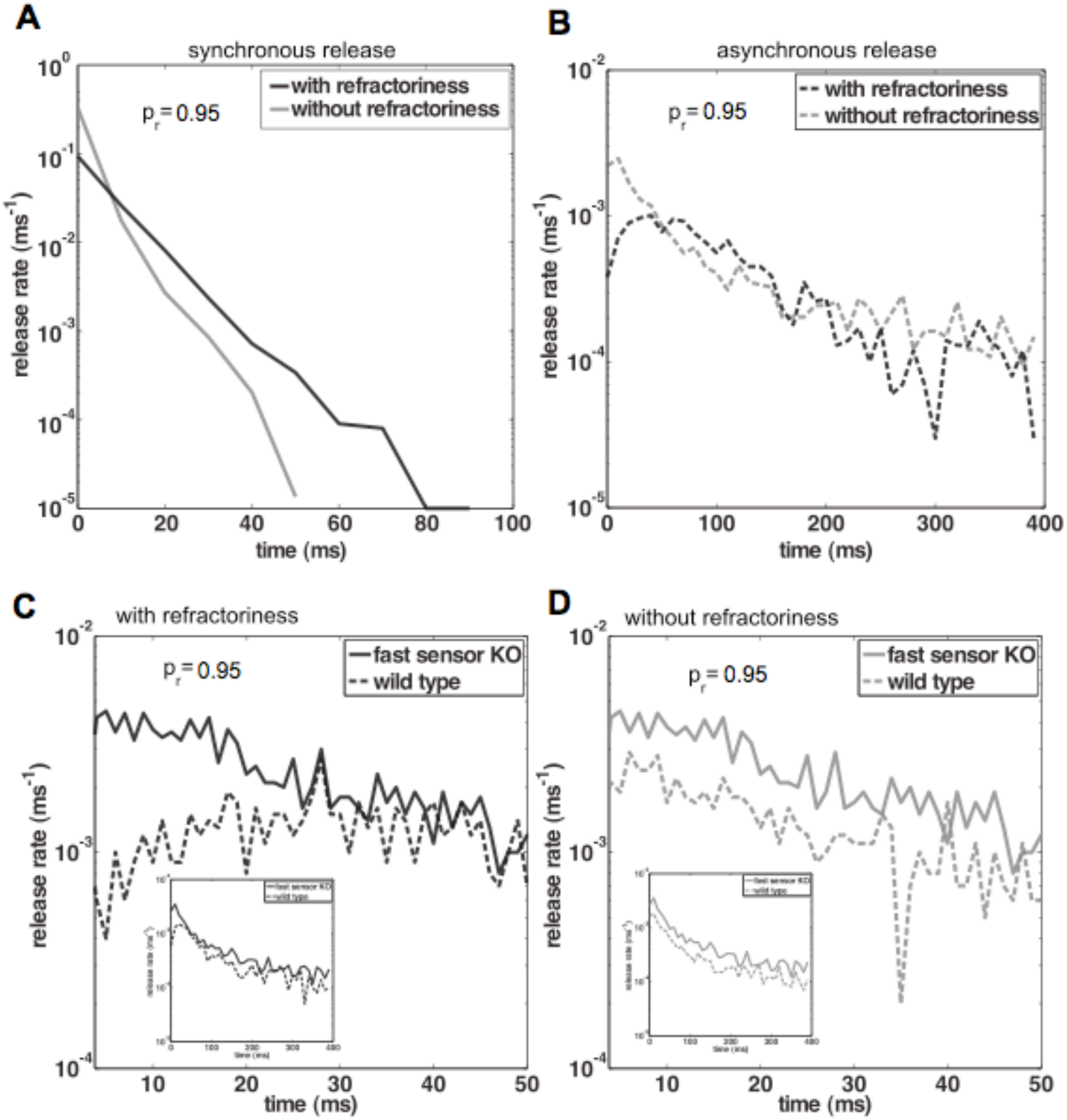

**Fig. 6 Components of synchronous and asynchronous release separated out for a synapse with and without refractoriness. (A)** For a synapse with refractoriness the synchronous release has a shorter, broader peak than the synapse without refractoriness. **(A)(B)** The asynchronous release channel encompasses more events for synapse with refractoriness compared to without refractoriness. Neurotransmitter release profile for fast sensor KO and wild type for a synapse with and without refractoriness. **(C)** The neurotransmitter release profiles for asynchronous release in wild type and fast sensor KO

varieties of the synapse with refractoriness (grey) diverge as they approach shorter time scales of less than 20 ms. Fast release through the synchronous pathway suppresses release from the asynchronous pathway due to the refractory period in the wild type, leading to a dip in asynchronous release. **(D)** The release profiles of wild type and fast sensor KO run almost parallel through the 400 ms transient in the synapse without (black) a refractory period. The transgenic fast sensor KO in both kinds of synapses (with and without refractoriness) is more elevated than the wild type as there is no depletion of vesicles, through the synchronous pathway, from the limited resource available in the RRP.

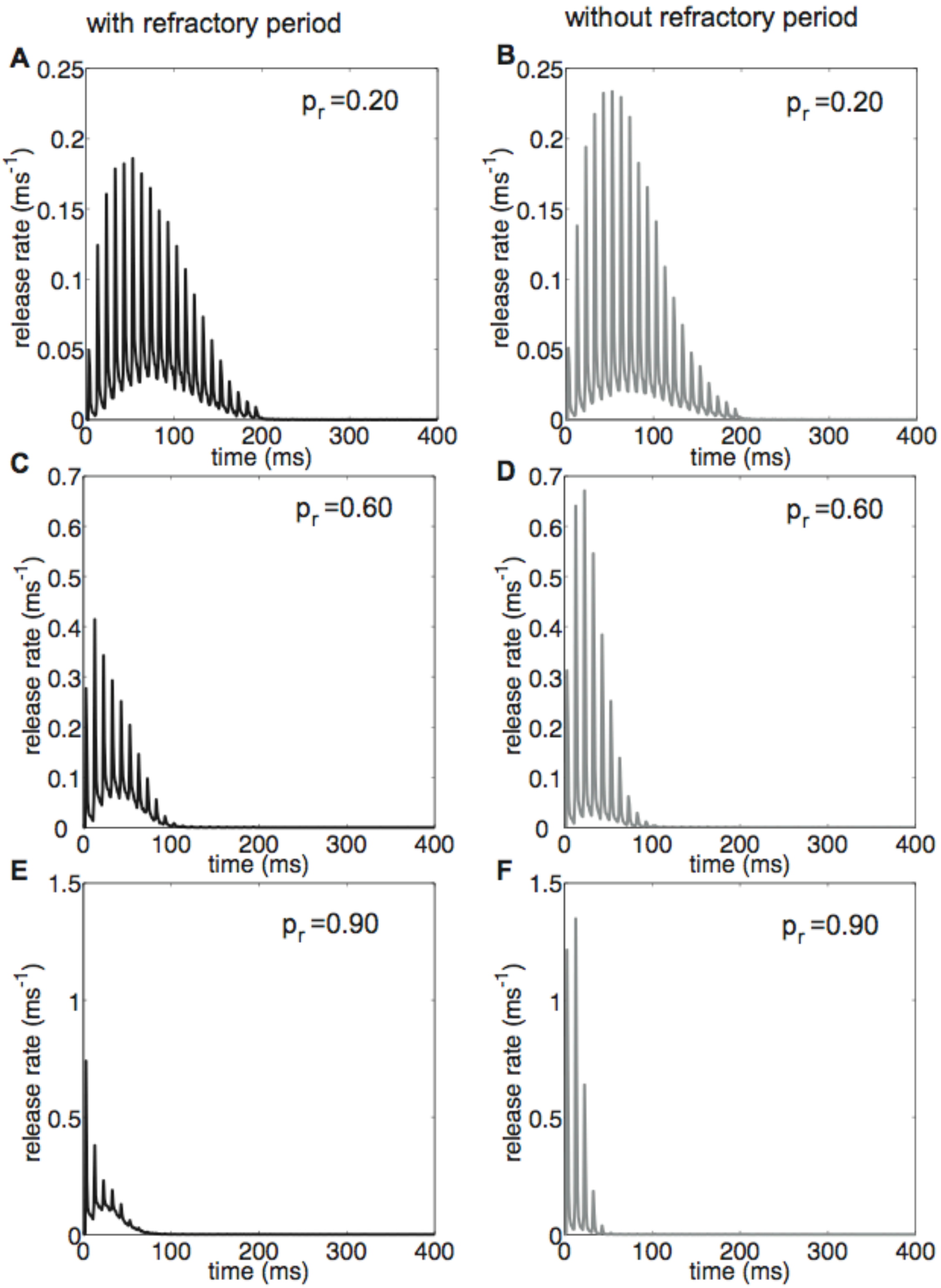

**Fig. 7 Response to a 200 ms at 50 Hz rate stimulus protocol administered to a model CA3-CA1 synapse with seven docked vesicles** The base level asynchronous release was higher in the synapse with refractoriness (gray) whereas the synapse without refractoriness (black) had higher peak release rates. The refractoriness allows the asynchronous release pathway to contribute more to the release. The rates of facilitation and depression were also characteristically different for these synapses.

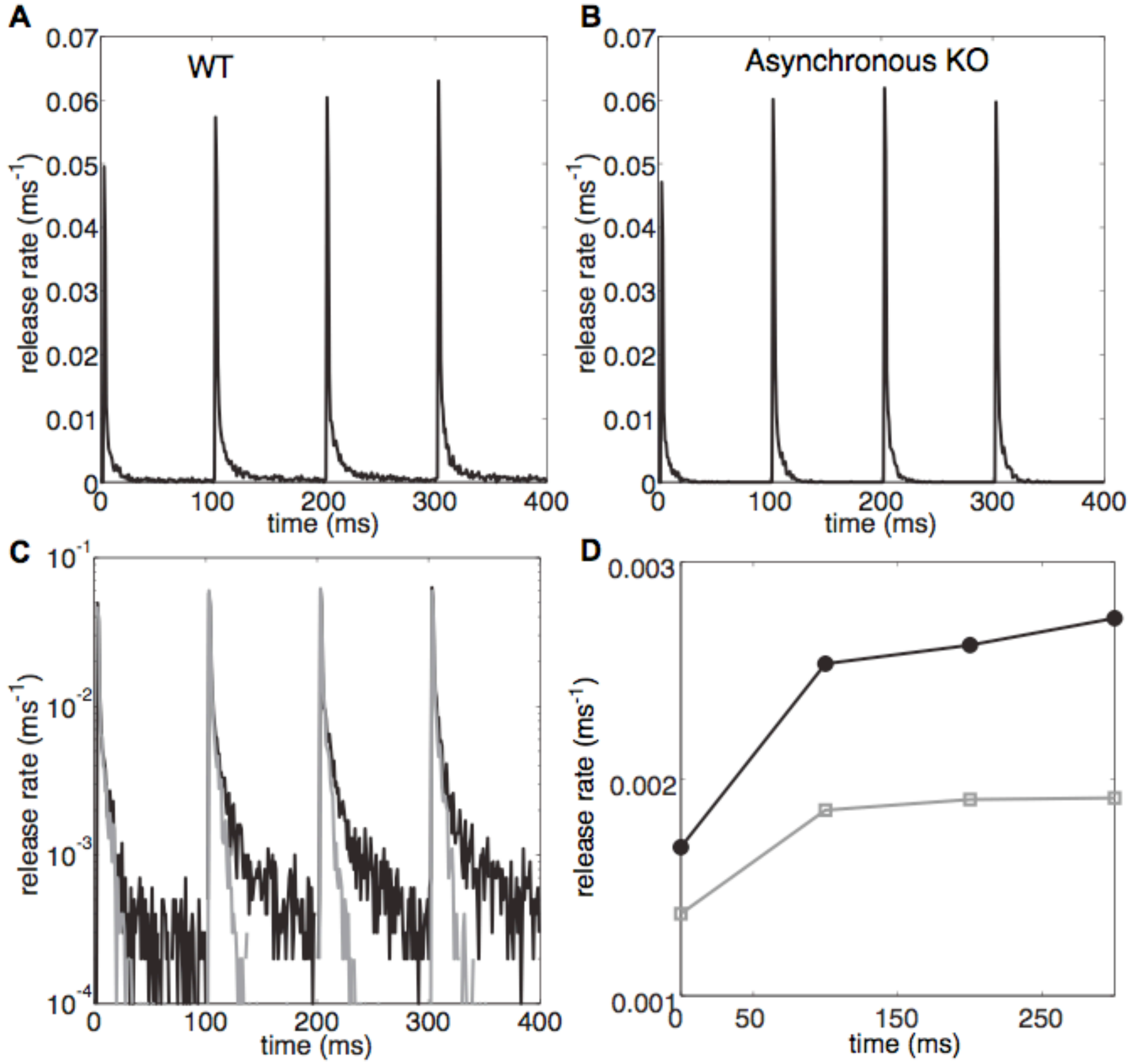

**Fig. 8 Response to 10 Hz train stimuli.** Release rate for wild type (**A**) and asynchronous release sensor KO (**B**) plotted in 1 ms bins. The same data is plotted on a log scale to show the elevated long tail of release (black line) due to the presence of asynchronous sensor in the wild type (**C**). The grey line in (**C**) is asynchronous sensor KO. In (**D**) total release rate (100 ms bins) for each stimuli is shown (wild type – black line, asynchronous KO – grey line).

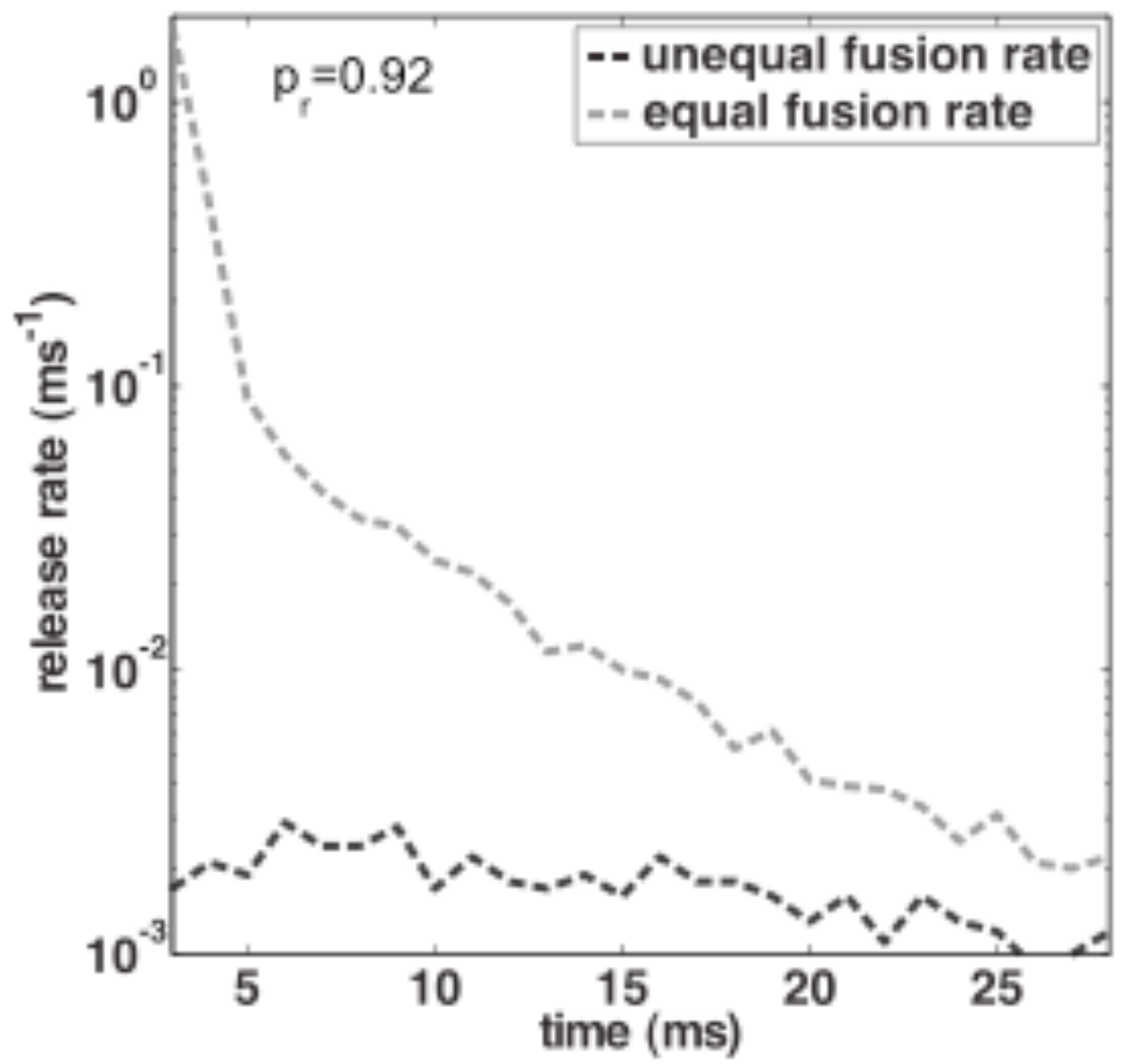

**Fig. 9 Release profile through the asynchronous pathway with identical vesicle fusion rates for synchronous and asynchronous release, compared with unequal fusion rates.** There is a sharp peak in the asynchronous release after the stimulus that coincides with the calcium signal at the active zone when the vesicle fusion rate is equal for the synchronous and asynchronous case. This peak seen in the simulations is not consistent with observed data. However, slowing down the fusion rate by a factor of 40 matches the data for spontaneous asynchronous release.

# Tables

**Table.1 Model Parameters**

| **Parameter [and reference]** | **Value** |
|---|---|
| Calcium diffusion Constant ($D_{Ca}$) [15] | 220 $\mu m^2/s$ |
| Calbindin diffusion constant (D_cb) [62] | 28 $\mu m^2/s$ |
| PMCA diffusion Constant ($D_{PMCA}$) | 0 $\mu m^2/s$ |
| Voltage dependent calcium channel (VGCC) diffusion constant ($D_{vgcc}$) | 0 $\mu m^2/s$ |
| Glutamate diffusion constant ($D_{glu}$) [63] | 200 $\mu m^2/s$ |
| Resting intracellular calcium concentration | 100 nM |
| Intracellular calbindin concentration [64] | 45 µM |
| PMCA surface density[65] | 180 $\mu m^{-2}$ |
| VDCC number [11] | 1 - 208 |
| Distance between the active zone and the VDCC cluster ($l_c$) [4] | 10 – 400 nm |
| Location of local $Ca^{2+}$ measurement | 10 nm (⊥ distance) from the active zone |
| Maximum radius of the VDCC cluster | 66 nm |
| **Calbindin-D28k [66]** | |
| Association rate, high affinity site ($kh_+$) | $0.55\times10^7\ M^{-1}\ s^{-1}$ |
| Dissociation rate, high affinity site ($kh_-$) | $2.6\ s^{-1}$ |
| Association rate, medium affinity ($km_+$) | $4.35\times10^7\ M^{-1}\ s^{-1}$ |
| Disassociation rate, medium affinity ($km_-$) | 35.8 $s^{-1}$ |
| **PMCA [65]** | |
| Association rate ($kpm_1$) | 1.5 ×$10^7$ $M^{-1}$ $s^{-1}$ |
| Disasociation rate ($kpm_2$) | 20 $s^{-1}$ |
| Transition rate 1 ($kpm_3$) | 20 $s^{-1}$ |
| Transition rate 2 ($kpm_4$) | 100 $s^{-1}$ |
| Leak rate ($kpm_{leak}$) | 12.5 $s^{-1}$ |
| VDCC [10] | $a_i(v) = a_{i0}\ \exp(v/v_i)$ and $b_i(v) = b_{i0}\exp(-v/v_i)$<br><br>Action potential transient reproduced from [10] |
| $\alpha_{10}, \alpha_{20}, \alpha_{30}, \alpha_{40}$ | 4.04, 6.70, 4.39, 17.33 $ms^{-1}$ |
| $\beta_{10}, \beta_{20}, \beta_{30}, \beta_{40}$ | 2.88, 6.30, 8.16, 1.84 $ms^{-1}$ |
| $v_1, v_2, v_3, v_4$ | 49.14, 42.08, 55.31, 26.55 mV |
| **Phenomenological Calcium sensor model for the entire active zone** | |
| Association rate, synchronous release ($ks_+$) | 1.91 ×$10^8$ $M^{-1}s^{-1}$ |
| Dissociation rate, synchronous release ($ks_-$) | 7.25 ×$10^3$ $s^{-1}$ |
| Association rate, asynchronous release ($ka_+$) | 3.68 ×$10^6$ $M^{-1}s^{-1}$ |
| Dissociation rate, asynchronous release ($ka_-$) | 26 $s^{-1}$ |
| b, $\gamma$, $\gamma^1$, $\varepsilon$ | 0.25, 6×$10^3$ /s, 0.417×$10^{-3}$ /s, 6.34 ms |
| **Discrete Calcium sensor model (Fig. 1)** | |
| Association rate, synchronous release ($ks_-$) | 1.91 ×$10^8$ $M^{-1}s^{-1}$ |
| Dissociation rate, synchronous release ($ks_-$) | $7.25 \times 10^3$ $s^{-1}$ |
| Association rate, asynchronous release ($ka_+$) | $3.68 \times 10^6$ $M^{-1}s^{-1}$ |
| Dissociation rate, asynchronous release ($ka_-$) | 26 $s^{-1}$ |
| b, γ, δ, ε, a | 0.25, $2\times10^3$ $s^{-1}$, $0.417\times10^{-3}$ $s^{-1}$, 6.34 ms, 0.025 |

# Supporting Information:

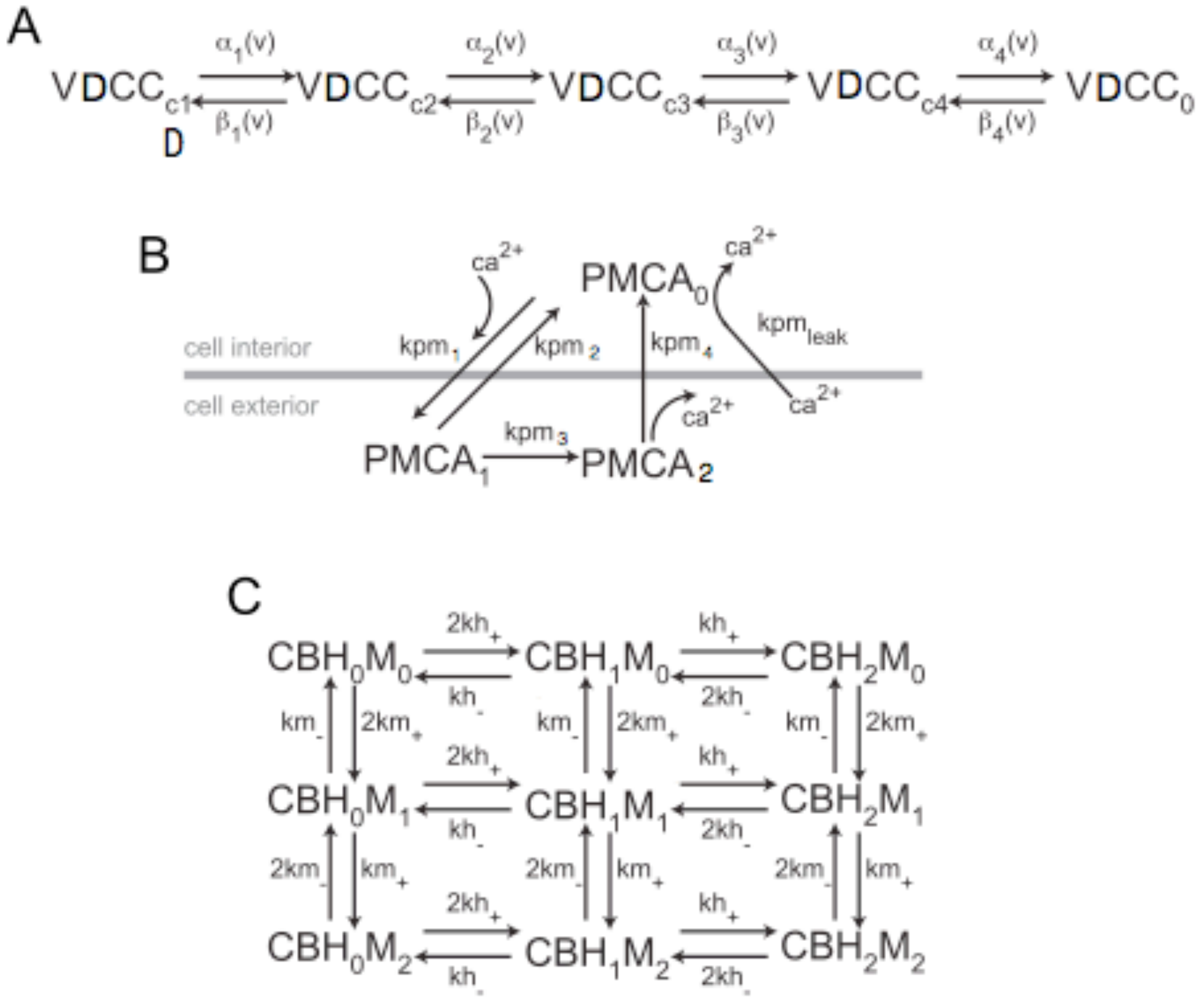

**Fig. S1 Kinetic schemes for Voltage Gated Calcium Channels, PMCA pumps and Calbindin.**

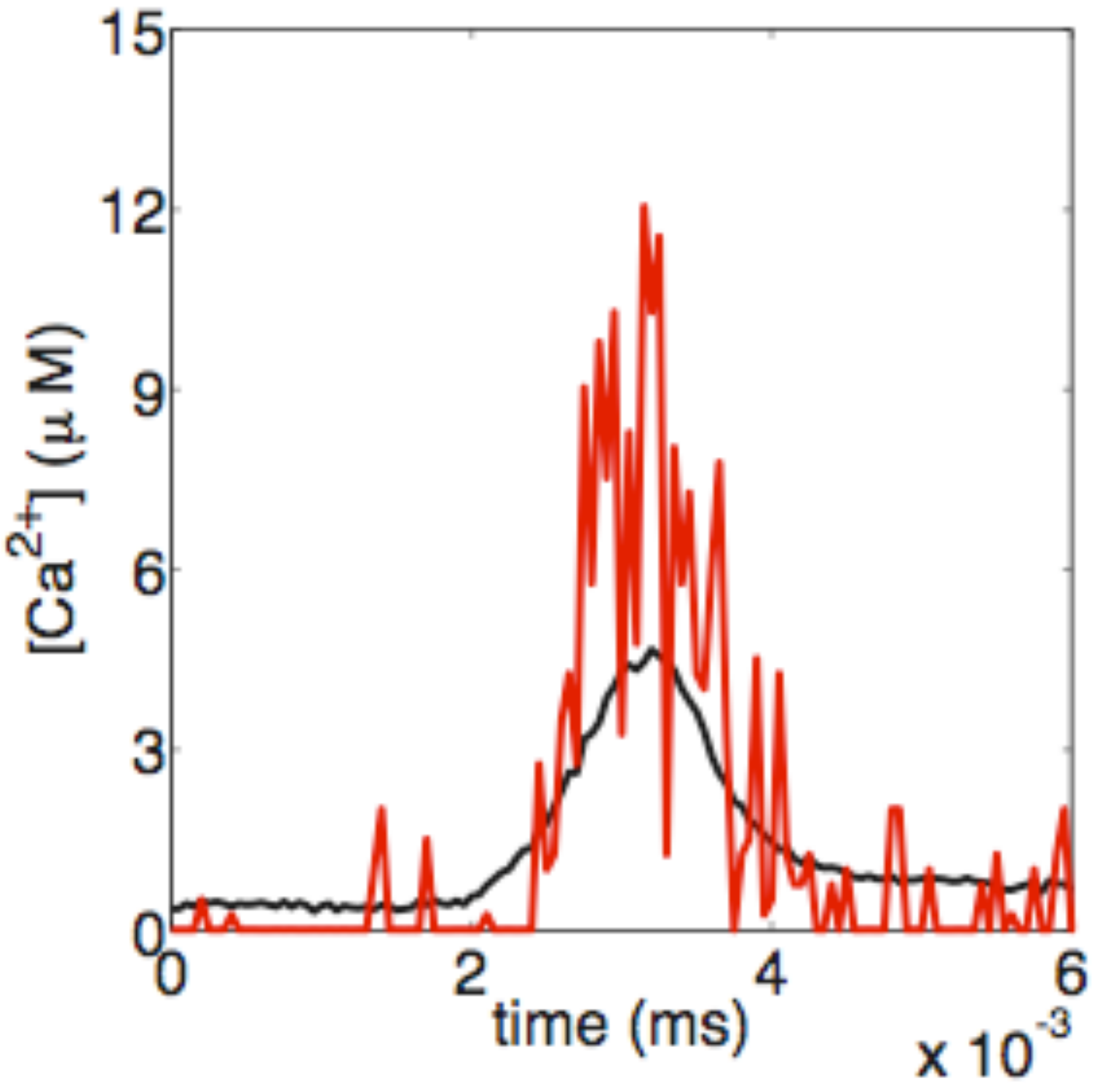

**Fig. S2 Sample calcium transient.** Single run (red) and average of 1000 trials (black) for release probability of 20% generated by 48 channels at 250 nm from the active zone. The calcium is measured 10 nm above the active zone.

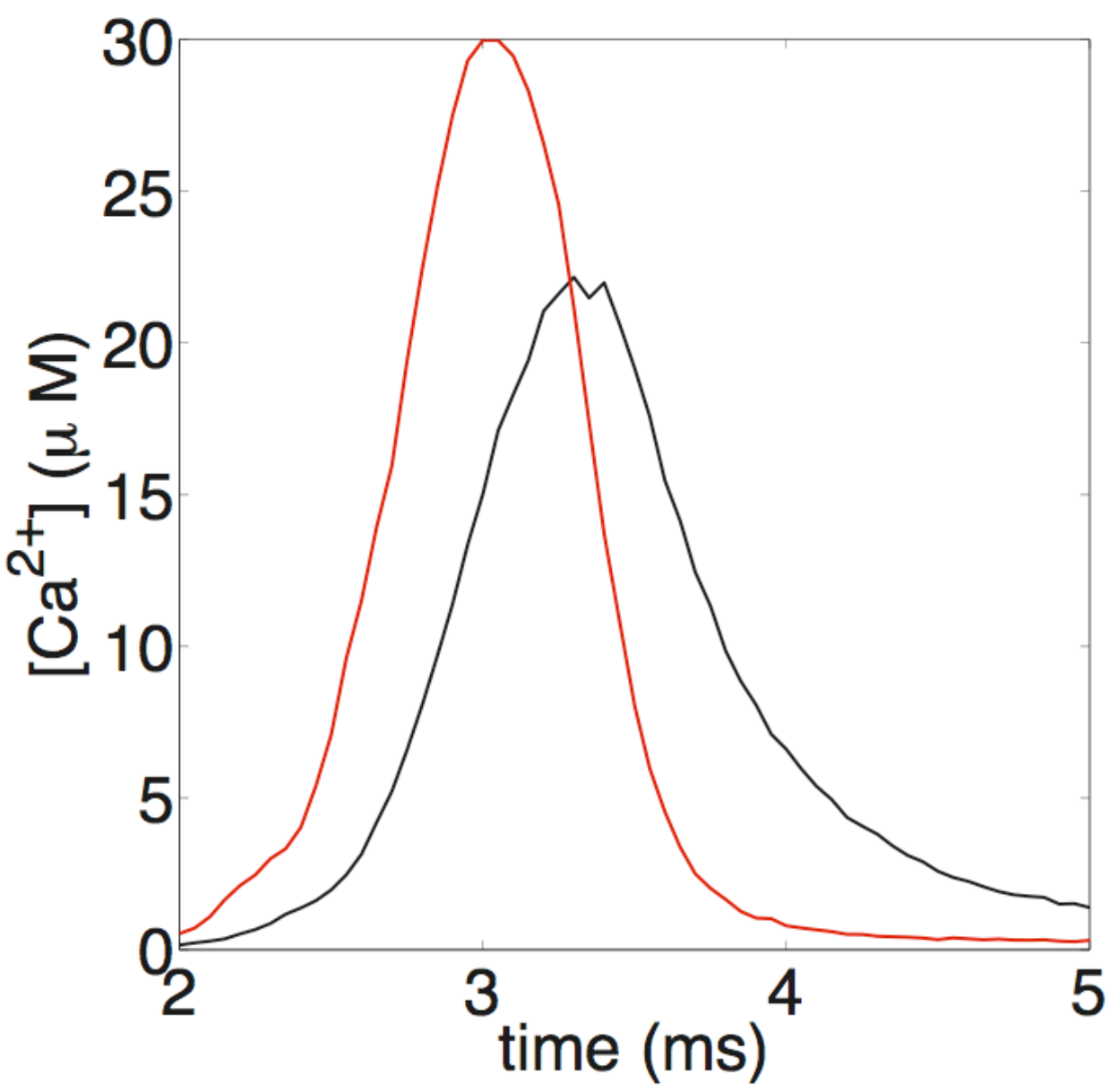

**Fig. S3 Sample local calcium transients for different distances.** The calcium transient generated by 32 VGCCs placed at 100 nm (red) and 160 VGCCs at 400 nm (black) lead to the same release probability of ~0.9. Ultrasynaptic structure such as the number and placement of the channels with respect to the calcium sensor modulate the shape of the calcium local calcium transient leading to non-overlapping neurotransmitter response curves for various $l_c$ seen in Fig. 2A.

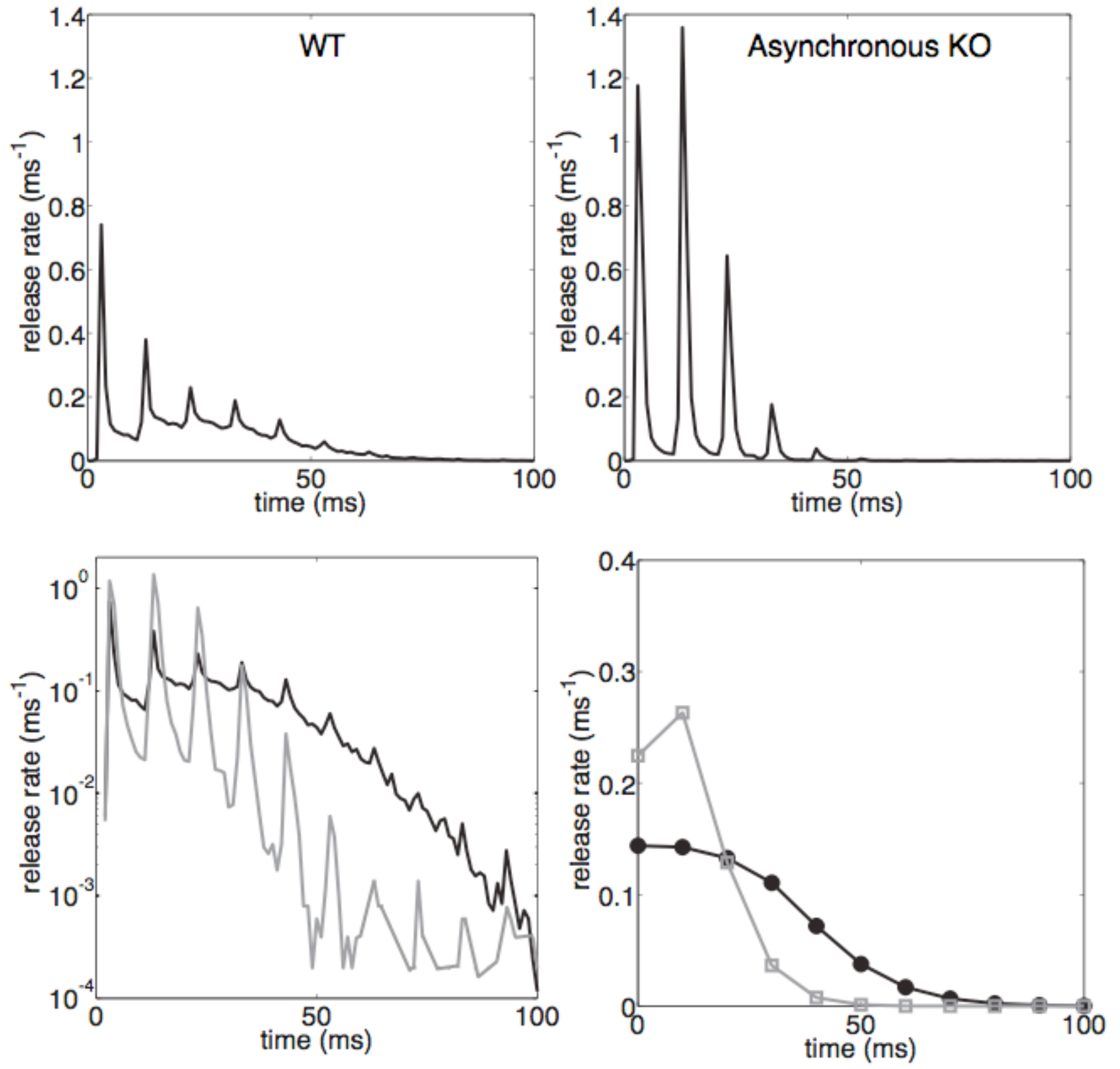

**Fig. S4 Response to 100 Hz train stimuli for a high release probability synapse ($p_r$=0.9).** Release rate for wild type (**A**) and asynchronous release sensor KO (**B**) plotted in 1 ms bins. The same data is plotted on a log scale to show higher base level release (black line) due to the presence of asynchronous sensor in the wild type (black line **C**). The grey line in (**C**) is asynchronous sensor KO. Here the effect of including the asynchronous sensor is completely opposite to simulations carried out for low frequency stimulus (10 Hz) in Fig. 8. For this fast stimulus the asynchronous release does not build up enough to benefit the subsequent incoming stimulus. The peak release rate for wild type (**A**) remains

lower than peak release rate of the KO. This is because the forward binding rate of asynchronous pathway is fast enough to compete for the incoming calcium ions but it is too slow to release (due to much slower fusion rate). This is also leads to overall reduction in release rate in the wild type as seen in (**D**). Data binned in 10 ms bins for each stimuli is shown here (wild type – black line, asynchronous KO – grey line).